\documentclass[a4paper, 12pt]{article} % use larger type; default would be 10pt

\usepackage[utf8]{inputenc} % set input encoding (not needed with XeLaTeX)

\usepackage{geometry} % to change the page dimensions
\usepackage{graphicx} % support the \includegraphics command and options
\usepackage{color} %Herb added
\usepackage[parfill]{parskip} % Activate to begin paragraphs with an empty line rather than an indent

\usepackage{amsmath} % Herb added for tfrac
\usepackage{booktabs} % for much better looking tables
\usepackage{array} % for better arrays (eg matrices) in maths
\usepackage{paralist} % very flexible & customisable lists (eg. enumerate/itemize, etc.)
\usepackage{verbatim} % adds environment for commenting out blocks of text & for better verbatim
\usepackage{subfig} % make it possible to include more than one captioned figure/table in a single float
\usepackage{empheq} %Herb added for emphasizing equations
\usepackage{epstopdf} %Herb added
\usepackage{multicol} %Herb added
\usepackage{float}

\usepackage{fancyhdr} % This should be set AFTER setting up the page geometry
\usepackage{sectsty}
\allsectionsfont{\sffamily\mdseries\upshape} % (See the fntguide.pdf for font help)
\usepackage[nottoc,notlof,notlot]{tocbibind} % Put the bibliography in the ToC
\usepackage[titles,subfigure]{tocloft} % Alter the style of the Table of Contents

\numberwithin{equation}{section} %Herbadded

\title{A description of the Thomas-Fermi ion with \\ fast converging function series}
\author{Herbert E. Müller \\ \normalsize http://herbert-mueller.info/}
\date{} 
\begin{document}
\maketitle

\begin{flushleft}

\begin{abstract}
This article concerns the description of the electron sea of an atomic ion with the Thomas-Fermi model. The normalized ion radius $X$, ionization potential $b$ and electronic binding energy $B$ of the Thomas-Fermi ion are functions of the ratio $N$ of electrons to protons in the ion. A scheme is given to calculate the Taylor series of $X(N)$, $b(N)$ and $B(N)$. With this scheme, the Taylor coefficients are calculated up to 5th order. The obtained 0th to 3rd order coefficients agree with the values presently available in the literature. To the authors knowledge, the 4th and 5th order coefficients are new results. It is then argued that a different series description of these functions, based on the Taylor series of $c(N) \equiv b^{-1/3}X^{-4/3}$, leads to a significant improvement in convergence. 
\end{abstract}

\newpage 

\tableofcontents

\newpage 

\section{Introduction}

\subsection{Elementary quantities}

Throughout this article we (the author and the readers) use the normalized units and most of the notation that are standard in texts about the Thomas-Fermi ion \cite{Englert}.  We abbreviate \textit{electron} by \textit{e}, and \textit{proton} by \textit{p}.\\ 
\quad
\begin{multicols}{2}
%First column
Number of protons: \textit{Z}

Unit of Length: $a_B/1.1295 Z^{1/3}$ 

Radius from the nucleus: $x$ 

$e$:$p$-ratio inside radius $x$: $n(x) $

%Second column
Numerical factor $1.1295 = \left( 128/9 \pi^2 \right)^{1/3}$

Unit of Energy: $2ZRy\cdot 1.1295 Z^{1/3}$

Ion radius: $X$

overall $e$:$p$-ratio: $n(X) \equiv N$
\end{multicols}

Potential energy of an $e$ due to the nucleus alone:  $-\frac{1}{x}$

Overall potential  energy of an $e$:  $ \phi \quad (<0) $

Ionization potential: $b \equiv -\phi(X)$

$e$:$p$ number density:  $\rho $

Fermi energy: $\rho^{2/3}$

Fermi surface: $\phi+\rho^{2/3} \quad (<0)$

Electronic binding energy \textit{per proton}:  $ B$

\subsection{Outline}

Our goal is to calculate the ion radius $X$, the ionization potential $b$ and the binding energy $B$  as functions of the $e$:$p$-ratio $N$. The steps to this goal are the following.

In section 1 we collect the basic formulas describing the Thomas-Fermi (TF) ion. 

In section 2 we express the Thomas-Fermi differential equation (TF DE) with the roles of $x$ and $\chi$ interchanged: the radius $x$ is the dependent variable, the screening  number $\chi$ is the independent variable. We solve the new DE in a semi-analytical manner for $x(\chi;a)$, where $a$ is the initial slope of the screening function $\chi(x)$. 

In section 3 we will then be able to express  $X$, $b$, $B$  and  $N$ as Taylor series in $a$. 

In section 4 we express $X$, $b$, $B$  and  $a$ as Taylor series in $N$. 

In section 5 we derive new series for $X(N)$, $b(N)$, $B(N)$  and  $a(N)$, with faster convergence. 

In section 6 we sum up and plot the results. 

\subsection{Description of the TF ion with the potential $\phi$}

The following basic equations of the Thomas-Fermi atom or ion can be found in most textbooks on classical quantum mechanics \cite{Englert}.

Let's start the description of the TF ion with some potential $\phi$, with  the constraints
\begin{equation}
\phi \rightarrow -1/x \quad (x \rightarrow 0) \qquad \qquad \phi= -bX/x \quad (x \geq X) \label{e1.1}
\end{equation}
Outside the ion radius X, $\phi$ is a Coulomb potential. $b$ is the ionization potential. From electrostatics, we have
\begin{equation}
b =(1-N)/X \label{e1.2}
\end{equation} 

The $e$:$p$ number density is given by Poisson's equation: 
\begin{equation}
\rho=-\Delta \left( \phi+\frac{1}{x} \right) \label{e1.3}
\end{equation} 
with $\rho(x)=0$ for $x \geq X$. 

By inserting this equation into 
\begin{equation*}
\frac{dn}{dx} = - x^2 \rho(x)
\end{equation*} 
and integrating we obtain the $e$:$p$-ratio inside radius $x$
\begin{equation}
n=-x^2\frac{d}{dx} \left( \phi+\frac{1}{x} \right) \label{e1.4}
\end{equation} 
with  $n(0)=0$ and $n(x) \equiv N$ for $x \geq X$. 

The equations given thus far allow us to construct physically reasonable state functions $\phi(x)$, $n(x)$ and $\rho(x)$, given the overall $e$:$p$-ratio $N$ and the ion radius $X$. We can then proceed to calculate the binding energy \textit{per proton}
\begin{equation}
B=\int_0^{X} dx x^2 \rho \left( \frac{1}{2x}-\frac{\phi}{2}-\frac{3}{5}\rho^{2/3} \right)  \label{e1.5}
\end{equation}

The actual state of the ion is the one with the highest binding energy $B$. To find this state, we vary the binding energy: 
\begin{equation*}
\delta B=-\int_0^{X} dx x^2\delta \rho \left(\phi+\rho^{2/3}\right)
\end{equation*}
while keeping the overall $e$:$p$ ratio constant:
\begin{equation*}
\delta N=\int_0^{X} dx x^2\delta \rho =0
\end{equation*}
Apparently, the maximal binding energy is achieved when the Fermi-surface is flat: 
\begin{equation*}
\phi+\rho^{2/3} = -b'
\end{equation*} 
The constant $b'$ on the RHS still depends on our choice of the ion radius $X$. However, it is intuitively clear that the ion radius is not a free parameter, but is determined by $N$. We therefore vary $X$ in the next step until the binding energy becomes minimal with respect to $X$. This will be achieved when the eletron density continuously approaches 0 as $x$ approaches $X$, i. e. we will have
\begin{equation*}
b'=b
\end{equation*}
In this way we obtain the relation 
\begin{equation}
\rho = \left(-\phi-b \right)^{3/2} \quad \label{e1.6}
\end{equation}

Incidentally, the last four unnumbered equations contain an important relation between the binding energy $B(N)$ and the ionization potential $b(N)$: 
\begin{equation*}
\delta B=b \delta N
\end{equation*}
If we drop the restriction $\delta N= 0$, the physical interpretation is straightforward: addition of an electron ($\delta N = 1/Z$) to an ion with a level Fermi surface increases the binding energy $ZB$ by the ionization potential $b$! It follows that the binding energy (now again per proton) is the primitive of the ionization potential with respect to the  $e$:$p$-ratio:
\begin{equation}
B(N)= \int_0 dN b(N) \quad \label{e1.7}
\end{equation}

\subsection{Description of the TF ion with the screening number $\chi$}

For mathematical convenience, we now describe the potential with the screening number $\chi$, according to 
\begin{equation}
\phi = -\frac{\chi}{x}-b \label{e1.8}
\end{equation}
The earlier statements about $\phi$ translate to 
\begin{equation}
\chi(0)=1 \qquad \qquad \chi(X)=0 \qquad \qquad \chi(x)=-b(x-X) \quad (x>X) \label{e1.9}
\end{equation}

The fraction of electrons inside radius \textit{x} is now 
\begin{equation}
n(x)= 1-\chi -x\psi \quad 
\end{equation}

Here we have introduced the slope of the screening function
\begin{equation}
\psi \equiv -\frac{d\chi}{dx} \quad \label{e1.11}
\end{equation}
The initial slope is customarily designated by $a$, and the final slope equals the ionization potential $b$ (see eqn \eqref{e1.9}):
\begin{equation}
\psi(0) \equiv a \qquad \qquad \psi(x)=b \quad (x \geq X)  \label{e1.12}
\end{equation}

Poisson's equation for the electron density \eqref{e1.3} is now
\begin{equation}
\rho=-\frac{1}{x}\frac{d\psi}{dx}=\frac{1}{x}\frac{d^2\chi}{dx^2} \label{e1.13}
\end{equation}

The binding energy \eqref{e1.5} becomes (after a short calculation, see appendix A) 
\begin{equation*}
B=\frac{a-b(1-N)}{2} - \int_0^X dx x \rho \left( \frac{3}{5}x \rho^{2/3} - \frac{\chi}{2}\right)
\end{equation*}
According to eqn \eqref{e1.6}  the maximum binding energy $B$ is obtained for 
\begin{equation}
\rho = \left( \frac{\chi}{x} \right)^{3/2} \label{e1.14}
\end{equation}

Inserting this density in Poisson's equation \eqref{e1.13}, we obtain the Thomas-Fermi differential equation (TF DE):
\begin{equation}
\frac{d^2\chi}{dx^2} = \frac{\chi ^{3/2}}{x^{1/2}} \label{e1.15}
\end{equation}

The solution of the TF DE will maximise the binding energy to
\begin{equation*}
B= \frac{a-b(1-N)}{2}-\frac{1}{10}\int_0^X dx x^{-1/2} \chi^{5/2}
\end{equation*}
In Appendix A it is shown that the last integral is $\frac{5}{7}(a-b(1-N))$. Thus the binding energy finally becomes 
\begin{equation}
B= \frac{3}{7}(a-b(1-N)) \label{e1.16}
\end{equation}

\subsection{Mathematical relations between the quantities of interest}

In deriving the TF DE, we have come across three interesting relations between the initial slope of the screening function $a$, the ionization potential = final slope of the screening function $b$, the ion radius $X$, the overall $e$:$p$-ratio $N$ and the binding energy $B$: 
\begin{itemize}
\item eqn. \eqref{e1.2}\enskip: $N=1-Xb$ 
\item   eqn. \eqref{e1.7}\enskip: $B=\int_0 dN b$
\item eqn. \eqref{e1.16}: $B= \frac{3}{7}(a-b(1-N))$
\end{itemize}
Any solution of the TF DE will give us a set of values $a$, $b$, $X$ from which we can calculate $B$ and $N$ by means of eqn.s \eqref{e1.2} and \eqref{e1.16}. 

Furthermore we can eke out an interesting differential equation between $a$, $b$ and $X$. The differentials of eqn.s \eqref{e1.16} and \eqref{e1.7} are $dB= \frac{3}{7}(da-db(1-N)+bdN)$ and $dB=bdN$. Setting them equal and rearranging, we obtain $4bdN=3(da-db(1-N))$. Eliminating $N$ with eqn \eqref{e1.2}, we obtain after a little calculation:
\begin{equation}
bXdb+4b^2dX+3da=0 \label{e1.17}
\end{equation}
Some more calculation gives 
\begin{equation}
(bX)^{7/3}dc=da \label{e1.18}
\end{equation}
where 
\begin{equation} 
c \equiv b^{-1/3}X^{-4/3}  \label{e1.19}
\end{equation} 
We will return to these equations later on. 

\section{General solution of the TF differential equation} 

The following general solution of the Thomas-Fermi differential equation is a Taylor-series in the parameter $a$ (initial slope of the screening function). The  coefficient-functions are calculated numerically.

\subsection{Changing the independent variable}

The second order TF DE can be split into two first order differential equations: 

\begin{minipage}{0.45\linewidth} 
\begin{equation} 
-\frac{d\chi}{dx} = \psi
\end{equation} 
\end{minipage} 
\begin{minipage}{0.45\linewidth} 
\begin{equation} 
-\frac{d\psi}{dx}= \frac{\chi^{3/2}}{x^{1/2}}
\end{equation} 
\end{minipage}

This system of DE's is non-linear. Historically, Riemann solved the non-linear DE for the plane sound wave by exchanging the roles of the dependent variable \textit{p} (pressure) and the independent variable \textit{x} (position) \cite{Sommerfeld}. We will use the same device here: $\chi$ becomes our independent variable, and \textit{x} and all other functions of interest become dependent variables. 

The two equations then must be written in the form    

\begin{minipage}[b]{0.45\linewidth} 
\begin{equation} 
\frac{dx}{d\chi} = -\frac{1}{\psi} \label{e2.9}
\end{equation} 
\end{minipage} 
\begin{minipage}[b]{0.45\linewidth} 
\begin{equation} 
\psi\frac{d\psi}{d\chi}= -\frac{\chi^{3/2}}{x^{1/2}} \label{e2.10}
\end{equation} 
\end{minipage}

Let us formally integrate these equations, starting at $\chi = 1$: 

\begin{minipage}[b]{0.45\linewidth} 
\begin{equation} 
x=\int^1\frac{d\chi}{\psi}  \label{e2.1}
\end{equation} 
\end{minipage} 
\begin{minipage}[b]{0.45\linewidth} 
\begin{equation} 
\psi^2 = a^2-2\int^1 d\chi \frac{\chi^{3/2}}{x^{1/2}}  \label{e2.2}
\end{equation} 
\end{minipage}

The form of eqn.s \eqref{e2.1} and \eqref{e2.2} suggests that the functions $x(\chi)$ and $\psi(\chi)$ can be calculated iteratively: the output of one equation is the input to the other. 

\subsection{Second normalization}

The two integral equations can be brought into a more symmetric form by substituting

\begin{minipage}[b]{0.45\linewidth} 
\begin{equation} 
\xi = \frac{ax}{1-\chi} 
\end{equation} 
\end{minipage} 
\begin{minipage}[b]{0.45\linewidth} 
\begin{equation} 
\eta =\left(\frac{\psi}{a}\right)^2 \\
\end{equation} 
\end{minipage}
\begin{minipage}[b]{0.45\linewidth} 
\begin{equation} 
t=\sqrt{1-\chi} \\
\end{equation} 
\end{minipage} 
\begin{minipage}[b]{0.45\linewidth} 
\begin{equation} 
K = \frac{2}{a^{3/2}}
\end{equation} 
\end{minipage}

The result is 

\begin{minipage}[b]{0.45\linewidth}
\begin{equation} 
\xi=\frac{2}{t^2} \int_0 dt \frac{t}{\eta^{1/2}}  \label{e2.3}
\end{equation} 
\end{minipage} 
\begin{minipage}[b]{0.45\linewidth} 
\begin{equation} 
\eta = 1-2K\int_0 dt \frac{(1-t^2)^{3/2}}{\xi^{1/2}}  \label{e2.4}
\end{equation} 
\end{minipage}

\pagebreak
\subsection{Taylor development in $K$}

The coupled integral eqn.s \eqref{e2.3} and \eqref{e2.4} contain the parameter $K = 2/a^{3/2}$. $K$ varies from 0 for the ``empty ion'' ($N = 0$) to 0.999367 for the neutral atom ($N = 1$). The fact that $K$ doesn't exceed 1 suggests that it can be used as an expansion coefficient for the functions $\eta(\chi)$ and $\xi(\chi)$. For $K=0$, i. e.  in lowest order, we have $\eta = 1$, or $\psi = a=constant$, which seems to be quite a reasonable starting point for the slope of the screening function, whatever the value of $N$. We therefore set
\begin{gather} 
\eta=1+K\eta_1+K^2\eta_2+K^3\eta_3+K^4\eta_4+K^5\eta_5 \ldots \\
\xi=1+K\xi_1+K^2\xi_2+K^3\xi_3+K^4\xi_4+K^5\xi_5 \ldots
\end{gather}

The integral eqn.s \eqref{e2.3} and \eqref{e2.4} also contain the powers $\eta^{-1/2}$ and $\xi^{-1/2}$. In fact, on several occasions we will have to calculate some power $\beta$ of a given power series. So let us deal with this problem first. 

Let $f(K)$ some Taylor series in $K$, starting with 1:
\begin{equation} 
f=1+Kf_1+K^2f_2+K^3f_3+K^4f_4+K^5f_5 \ldots \label{e2.8}
\end{equation} 
Raising this series to the power of $\beta$ gives 
\begin{equation} 
f^{\beta}=1+Kf_1^{(\beta)}+K^2f_2^{(\beta)}+K^3f_3^{(\beta)}+K^4f_4^{(\beta)}+K^5f_5^{(\beta)} \ldots 
\end{equation} 
The coefficients $f_m^{(\beta)}$ can be calculated by repeated application of the binomial theorem. The coefficients $f_1 \ldots f_5$ are:
\begin{align} 
f_1^{(\beta)}=&\beta f_1 \nonumber \\ 
f_2^{(\beta)}=&\beta f_2+\frac{\beta(\beta-1)}{2}f_1^2 \nonumber \\ 
f_3^{(\beta)}=&\beta f_3+\beta(\beta-1)f_1f_2+\frac{\beta(\beta-1)(\beta-2)}{6}f_1^2 \label{e2.5} \\
f_4^{(\beta)}=&\beta f_4+\frac{\beta(\beta-1)}{2}(f_2^2+2f_1f_3)+\frac{\beta(\beta-1)(\beta-2)}{2}f_1^2f_2 +\frac{\beta\ldots(\beta-3)}{24}f_1^4 \nonumber \\
f_5^{(\beta)}=&\beta f_5+\beta(\beta-1)(f_2f_3+f_1f_4)+ \frac{\beta(\beta-1)(\beta-2)}{2}(f_1^2f_3+f_1f_2^2) + \nonumber \\
&\frac{\beta\ldots(\beta-3)}{6}f_1^3f_2+ \frac{\beta\ldots(\beta-4)}{120}f_1^5 \nonumber 
\end{align} 

The power series for $\eta^{-1/2}$ and $\xi^{-1/2}$ in eqn.s \eqref{e2.3} and \eqref{e2.4}  are now formally 
\begin{gather} 
\eta^{-1/2}=1+K\eta_1^{(-1/2)}+K^2\eta_2^{(-1/2)}+K^3\eta_3^{(-1/2)}+
K^4\eta_4^{(-1/2)}+K^5\eta_5^{(-1/2)} \ldots \\
\xi^{-1/2}=1+K\xi_1^{(-1/2)}+K^2\xi_2^{(-1/2)}+K^3\xi_3^{(-1/2)}+
K^4\xi_4^{(-1/2)}+K^5\xi_5^{(-1/2)} \ldots 
\end{gather}
The coefficients $\eta_m^{(-1/2)}$ and $\xi_m^{(-1/2)}$ can be calculated with eqn \eqref{e2.5}.

After ordering in powers of $K$, eqn.s \eqref{e2.3} and \eqref{e2.4} can be rewritten as

\begin{minipage}[b]{0.45\linewidth}
\begin{equation} 
\xi_m=\frac{2}{t^2} \int_0 dt \: t \eta_m^{(-1/2)} \label{e2.6}
\end{equation} 
\end{minipage} 
\begin{minipage}[b]{0.45\linewidth} 
\begin{equation} 
\eta_{m+1} = -2\int_0 dt \: (1-t^2)^{3/2}\xi_m^{(-1/2)} \label{e2.7}
\end{equation} 
\end{minipage}
 
\subsection{Numerical integration}

The integration of eqn.s \eqref{e2.6} and \eqref{e2.7}  can be performed sequentially on the computer. We begin with $\eta_0 = 1$ $\rightarrow \eta_0^{(-1/2)} = 1$ $\rightarrow \xi_0 = 1$  $\rightarrow \xi_0^{(-1/2)} = 1$. The first non-trivial term $\eta_1$ is obtained by inserting $\xi_0^{-1/2} = 1$ into eqn \eqref{e2.7} . 

The result of the numerical integration of $\eta_m(t)$ and $\xi_m(t)$ is shown in figures (1) and (2). The color of the orders is red (1), yellow (2), green (3), cyan (4), blue (5), magenta (6).
It is striking that the $\eta_m(t)$ get small very quickly with increasing $m$. On the other hand the $\xi_m(t)$ vanish only slowly, since the radius $\xi(1)=aX$ of the neutral TF atom ($a=1.588071$ and $K=0.999367$) is infinite.
\begin{figure}[ht]
\begin{minipage}[b]{0.47 \linewidth}
\centering
\includegraphics[trim = 0pt 0pt 20pt 0pt, scale=1 ]{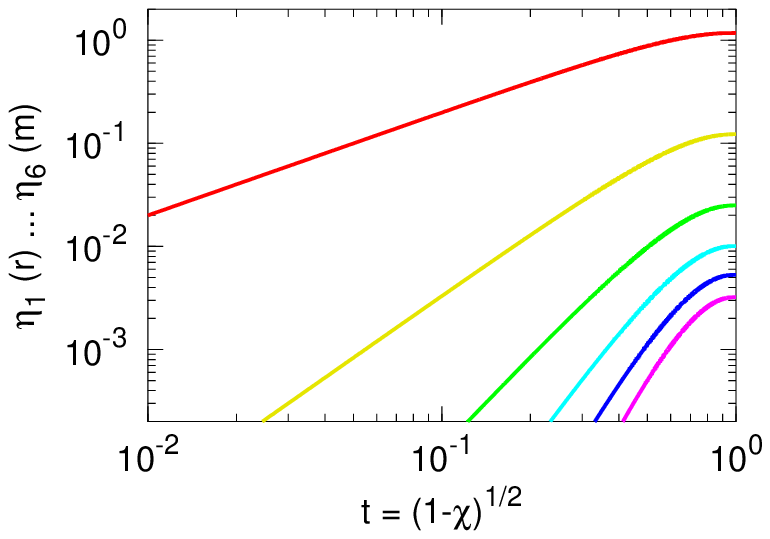}
\caption{Taylor coefficients of \\ $\eta(t,K)=1+K\eta_1(t)+K^2\eta_2(t)+\ldots$}
\end{minipage}
\hspace{0.04 \linewidth}
\begin{minipage}[b]{0.47 \linewidth}
\centering
\includegraphics[trim = 0pt 0pt 20pt 0pt, scale=1]{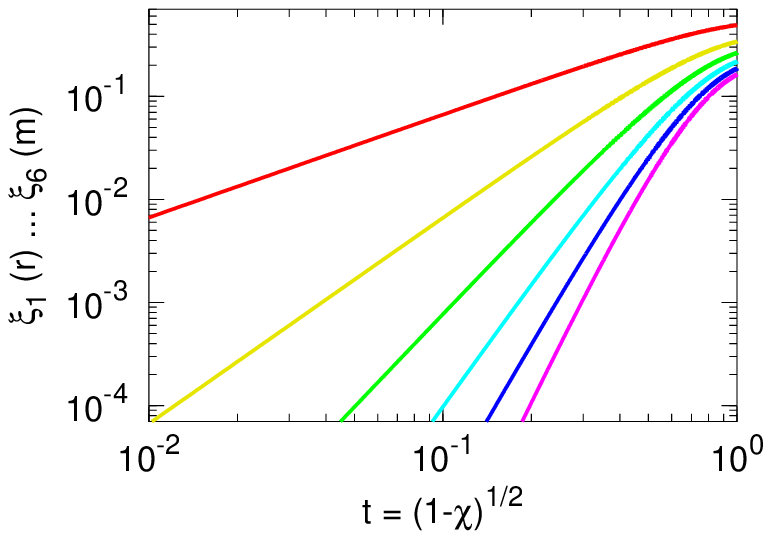}
\caption{Taylor coefficients of \\ $\xi(t,K)=1+K\xi_1(t)+K^2\xi_2(t)+\ldots$}
\end{minipage}
\end{figure}

The first impression is confirmed if we plot the partial sums of $x=\xi(1-\chi)/a$ and $\eta=(\psi/a)^2$ versus $\chi=1-t^2$ for the neutral atom and compare them to the known limiting functions. The approximation of $\eta(\chi)$ is fair, the approximation of $\chi(x)$ is poor.
\begin{figure}[ht]
\begin{minipage}[b]{0.47 \linewidth}
\centering
\includegraphics[trim = 0pt 0pt 20pt 0pt, scale=1 ]{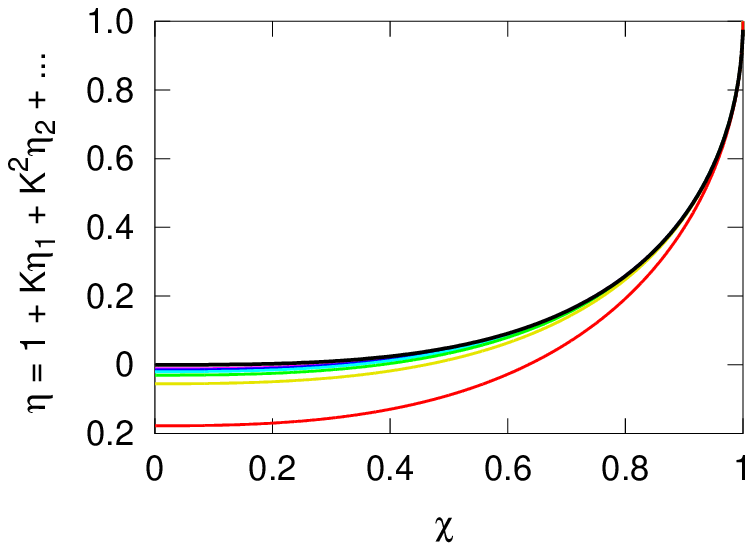}
\caption{Successive approximations of $\eta(\chi,K)=1+K\eta_1(\chi)+K^2\eta_2(\chi)+\ldots$ for $K=0.999367$ (neutral atom)}
\end{minipage}
\hspace{0.04 \linewidth}
\begin{minipage}[b]{0.47 \linewidth}
\centering
\includegraphics[trim = 0pt 0pt 20pt 0pt, scale=1]{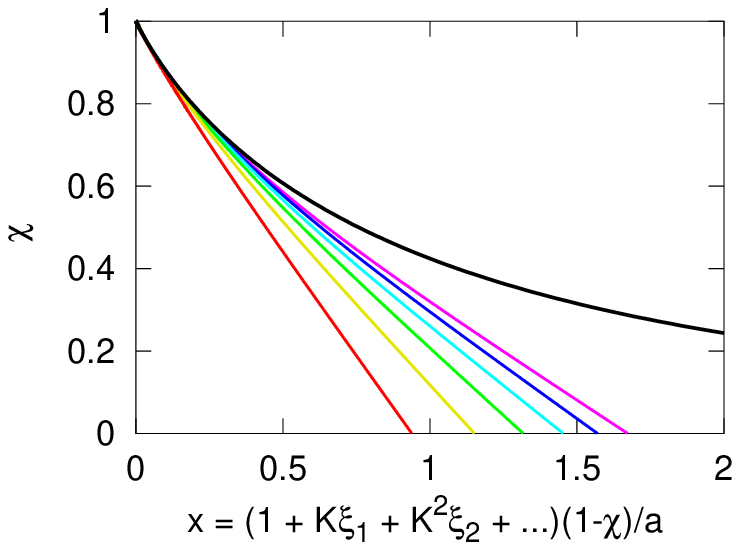}
\caption{Successive approximations of $x(\chi,K)=(1+K\xi_1(\chi)+K^2\xi_2(\chi)+\ldots)\times \\ (1-\chi)/a$ for $K=0.999367$ (neutral atom)}
\end{minipage}
\end{figure}

\section{Taylor series for $X^{-1}(K)$, $b(K)$, $B(K)$ and $N(K)$}

\subsection{Formulas for Taylor coefficients}

The integration of eqn.s \eqref{e2.6} and \eqref{e2.7} provides two of the Taylor series in $K$ we are looking for (remember $K=2/a^{3/2}$): 
\begin{align} 
aX = \xi(1) &= 1+K\xi_1(1)+K^2\xi_2(1)+K^3\xi_3(1)+K^4\xi_4(1)+K^5\xi_5(1) \ldots \label{e3.3} \\
(b/a)^2 =  \eta(1) &= 1+K\eta_1(1)+K^2\eta_2(1)+K^3\eta_3(1)+K^4\eta_4(1)+K^5\eta_5(1) \ldots \label{e3.4}
\end{align}
For physical reasons, we are interested in the power series for $b/a$ rather than $(b/a)^2$. From a mathematical point of view, the power series for $(aX)^{-1}$ has some advantages over the one for $aX$ ($(aX)^{-1} \rightarrow 0$ for $K \rightarrow 0.999367$, and eqn \eqref{e1.2}). These two series are 
\begin{align} 
b/a=\eta^{1/2}(1) & =1+K \eta_1^{(1/2)}+K^2 \eta_2^{(1/2)} +K^3 \eta_3^{(1/2)} +K^4 \eta_4^{(1/2)}+K^5 \eta_5^{(1/2)} \ldots \label{e3.1} \\
(aX)^{-1}=\xi^{-1}(1) & =1+K \eta^{(-1)}_1+K^2 \eta_2^{(-1)} +K^3 \eta_3^{(-1)} +K^4 \eta_4^{(-1)}+K^5 \eta_5^{(-1)} \ldots \label{e3.2}
\end{align} 
The RHSs of eqn.s \eqref{e3.1} and \eqref{e3.2} must be evalated at $t=1$.  The coefficients $\eta_m^{(1/2)}$ and $\xi_m^{(-1)}$ can be calculated with eqn \eqref{e2.5}. 
 
The Taylor series for the other two quantities we seek, $B(K)$ and $N(K)$, are obtained as follows. 

Eqn. \eqref{e1.16} for $B$ can be reformulated in terms of our new variables $\eta$ and $\xi$:  
\begin{equation} 
\frac{B}{a}=\frac{3}{7}(1-\eta(1)\xi(1))
\end{equation}
Inserting the series \eqref{e3.1} and \eqref{e3.2}, we obtain
\begin{align} 
\frac{B}{a}= -\frac{3}{7} & \left[    K(\eta_1+\xi_1)+K^2(\eta_2+\eta_1\xi_1+\xi_2)+K^3(\eta_3+\eta_2\xi_1+\eta_1\xi_2+\xi_3) + \right. \label{e3.5} \\
& \ \left.  K^4(\eta_4+\eta_3\xi_1+\eta_2\xi_2+\eta_1\xi_3+\xi_4)+ K^5(\eta_5+\eta_4\xi_1+\eta_3\xi_2+\eta_2\xi_3+\eta_1\xi_4+\xi_5) \ldots  \right] \nonumber
\end{align} 
The RHSs of eqn \eqref{e3.5} must be evalated at $t=1$.

Eqn. \eqref{e1.2} for $N$ now becomes
\begin{equation} 
N = 1-\xi(1)\eta^{1/2}(1) 
\end{equation}
Inserting the power series \eqref{e3.3} and \eqref{e3.1}, we obtain
\begin{align} 
N= - & \left[   K (\eta_1^{(1/2)}+\xi)+K^2 (\eta_2^{(1/2)}+\eta_1^{(1/2)}\xi_1+\xi_2) +K^3 (\eta_3^{(1/2)}+\eta_2^{(1/2)}\xi_1+\eta_1^{(1/2)}\xi_2+\xi_3)+\right. \nonumber \\
& \ \left. K^4 (\eta_4^{(1/2)}+\eta_3^{(1/2)}\xi_1+\eta_2^{(1/2)}\xi_2+\eta_1^{(1/2)}\xi_3+\xi_4 )+\right. \nonumber \\
& \ \left. K^5 (\eta_5^{(1/2)}+\eta_4^{(1/2)}\xi_1+\eta_3^{(1/2)}\xi_2+
\eta_2^{(1/2)}\xi_3+\eta_1^{(1/2)}\xi_4+\xi_5) \ldots  \right] \label{e3.6}
\end{align}
Again, the RHSs of eqn \eqref{e3.6} must be evalated at $t=1$.

\subsection{Taylor coefficients }

Let us collect the Taylor coefficients of $X^{-1}(K)$, $b(K)$, $B(K)$, $N(K)$ (and of two more quantities) in a table: 

\begin{table}[H]
\begin{tabular}{ |l|r|r|r|r|r|r|}
\hline
$f$ & $aX=\xi(1)$ & $(aX)^{-1}$ & $(b/a)^2=\eta(1)$ & $b/a$ & $B/a$ & $N$ \\ 
\hline
$f_0$ & 1& 1 & 1& 1 & 0 & 0 \\
$f_1$ & 0.490873 & $-$0.490873 & $-$1.178097 & $-$0.589049 & 0.294524 & 0.098175 \\
$f_2$ & 0.339148 & $-$0.098191 & 0.122481 & $-$0.112248 & 0.050000 & 0.062248 \\
$f_3$ & 0.263353 & $-$0.048674 & 0.024990 & $-$0.053624 & 0.021892 & 0.045145 \\
$f_4$ & 0.217190 & $-$0.030722 & 0.010085 & $-$0.032845 & 0.012502 & 0.035173 \\
$f_5$ & 0.185856 & $-$0.021795 & 0.005303 & $-$0.022715 & 0.008155 & 0.028663 \\
$f_6$ & 0.163069 & $-$0.016574 & 0.003220 & $-$0.016894 & 0.005768 & 0.024093 \\
\hline
\end{tabular}
\caption{Coefficients of the Taylor series $f=f_0+Kf_1+K^2f_2+K^3f_3+\ldots $ for selected functions $f$, with $K=(2/a)^{3/2}$.}
\end{table}

\subsection{Convergence}

We already know the values of all the quantities $f$  in table 1 for the neutral atom ($K=0.99936725$, the most difficult case to calculate). Let us see how quickly the series converge towards these values. 

\begin{table}[H]
\begin{tabular}{|l|r|r|r|r|r|r|r|} 
\hline
$f$ & $aX=\xi(1)$ & $(aX)^{-1}$ & $(b/a)^2=\eta(1)$ & $b/a$ & $B/a$ & $N$ \\ 
\hline
$S_0$ & 1 & 1 & 1 & 1 & 0 &  0 \\
$\ldots$ & \ldots & \ldots & \ldots & \ldots & \ldots &  \ldots \\
$S_3$ & 2.092136 & 0.362787 & $-$0.030081 & 0.245695 & 0.366125 & 0.205342 \\
$S_4$ & 2.308777 & 0.332142 & $-$0.020022 & 0.212933 & 0.378597 & 0.240426 \\
$S_5$ & 2.494047 & 0.310415 & $-$0.014735 & 0.190290 & 0.386726 & 0.269000 \\
$S_6$ & 2.656499 & 0.293903 & $-$0.011527 & 0.173460 & 0.392473 & 0.293002 \\
\hline
$S_\infty$ & $\infty$ & 0 & 0 & 0 & 0.428571 & 1 \\
\hline
\end{tabular}
\caption{Partials sums  $S_m = f_0+Kf_1+K^2f_2+\ldots+K^m f_m$ for $K=(2/a)^{3/2}=0.999367$ (neutral atom) and selected functions $f$.}
\end{table}

The convergence is moderate for $b^2$, $b$ and $B$, slow for $X$ and $X^{-1}$, and very slow for $N$. This is a problem. There is no use calculating a precise value of, say, the binding energy $B$ for some value of $K$, if we don't know the precise $e$:$p$-ratio $N$ for this $K$. 

Well, let's just proceed to the next point on our agenda, which is eliminiating the intermediate parameter $K$. When this is done, we will evaluate the convergence of the new series for $f(N)$. 

\section{Taylor series for $X^{-1}(N)$, $b(N)$, $B(N)$ and $a(N)$}
 
\subsection{Eliminating the parameter $K$}

Our problem is as follows. We are given the Taylor developments of two functions $N(K)$, see table 1, and $f(K)$, which could be any of the other quantities in table 1:   
\begin{gather} 
N=N_1K+N_2K^2+N_3K^3+N_4K^4+N_5K^5 \ldots \label{e4.1} \\
f=(K/2)^\alpha(f_0+f_1K+f_2K^2+f_3K^3+f_4K^4+f_5K^5 \ldots) \label{e4.4}
\end{gather}
The initial factor $(K/2)^\alpha=a^{-3\alpha/2}$ has been absorbed in $f$ in the course of the second normalization ($aX$, $b/a$, $B/a$ \ldots). In \eqref{e4.4} we have reverted to the more familiar "first normalization"  ($X$, $b$, $B$ \ldots).

We want to express $f$ as a power series in $N$. It is not difficult to see that this series must be of the following form:
\begin{equation} 
f=N^\alpha ( \tilde f_0+ \tilde f_1 N+ \tilde f_2 N^2+ \tilde f_3 N^3+ \tilde f_4 N^4+ \tilde f_5 N^5 \ldots)
\end{equation} 

We want to express the new coefficients $\tilde f_n$ as linear combinatons of the old coefficients $f_m$. 
To this end we first rewrite eqn \eqref{e4.1} as 
\begin{equation} 
\frac{N}{N_1}=K(1+Kh_1+K^2h_2+K^3h_3+K^4h_4+K^5h_5 \ldots) \label{e4.2}
\end{equation} 
where
\begin{equation} 
h_1=\frac{N_2}{N_1} \qquad h_2=\frac{N_3}{N_1} \qquad \ldots 
\end{equation} 

Elevating eqn \eqref{e4.2} to the power of $\beta$ gives 
\begin{equation} 
\left(\frac{N}{N_1}\right)^{\beta}=K^{\beta}(1+h_1^{(\beta)}K+h_2^{(\beta)}K^2+h_3^{(\beta)}K^3+h_4^{(\beta)}K^4+h_5^{(\beta)}K^5 \ldots) \label{e4.3}
\end{equation} 
with the coefficients $h_n^{(\beta)}$ given earlier in eqn \eqref{e2.5}.  
 
The $\tilde f_n$ are now obtained by repeatedly subtracting eqn \eqref{e4.3} from eqn \eqref{e4.1}, in such a way that the powers of $K$ on the RHS of \eqref{e4.1} are one by one eliminated. In the course of this elimination procedure, it appears that it is useful to introduce intermediate coefficients $g_n$ defined by 
\begin{equation} 
\tilde f_n=\frac{g_n}{2^{\alpha}N_1^{\alpha+n}} \quad (n \geq 0) \label{e4.6}
\end{equation} 

The Taylor series \eqref{e4.4} written with the $g_n$ is
\begin{equation} 
f=\left(\frac{N}{2N_1}\right)^\alpha\left\{ g_0 +g_1\frac{N}{N_1} +g_2\left(\frac{N}{N_1}\right)^2 +g_3 \left(\frac{N}{N_1}\right)^3 +g_4 \left(\frac{N}{N_1}\right)^4 + \ldots \right\} \label{e4.5}
\end{equation} 

The elimination procedure can now be formalized by the following scheme. It shows how to calculate the coefficients $g_n$ out of the known $f_m$ and $h_m^{(\beta)}$:
\begin{equation} 
\begin{array}{l l l l}
G_{0,0} = f_0  & G_{0,1} = f_1  & G_{0,2}=f_2 & G_{0,3}=f_3\\
G_{1,0} = G_{0,1}-G_{0,0}h_1^{(\alpha)}  & G_{1,1} = G_{0,2}-G_{0,0}h_2^{(\alpha)}  &
G_{1,2}=G_{0,3}-G_{0,0}h_3^{(\alpha)} & \ldots \\
G_{2,0} = G_{1,1}-G_{1,0}h_1^{(\alpha+1)} & G_{2,1} = G_{1,2}-G_{1,0}h_2^{(\alpha+1)} & \ldots &    \\
G_{3,0} = G_{2,1}-G_{2,0}h_1^{(\alpha+2)} & \ldots  & &  \\
\ldots & & &\\
\end{array} 
\end{equation} 

The general recursion relation in this scheme is 
\[G_{n,k} = G_{n-1,k+1}-G_{n-1,0}h_{k+1}^{(\alpha+n-1)} \]

Our $g_n$ are the first column of $G_{n,k} $:
\begin{equation} 
g_n=G_{n,0} 
\end{equation} 

\subsection{A linear transformation of Taylor coefficients}

The relation between the coefficients $f_m$ in the $K$-series \eqref{e2.8} and $g_n$ in the $N$-series \eqref{e4.5} is linear, i. e. it can be represented by a transformation matrix $T_{mn}(\alpha)$: 
\begin{equation} 
g_n=f_mT_{mn} \label{e4.7}
\end{equation} 

The $T_{mn}(\alpha)$ can be obtained with the scheme described above, by setting: 
\begin{equation} 
f_{m'}=\delta_{mm'} \Rightarrow T_{mn}=g_n
\end{equation} 

In the following subsection, we will mainly need the transformation matrix for $\alpha=-2/3$. A numerical evalation of the above scheme on the computer gives 

\[\mathbf{T} \left( \alpha=-\frac{2}{3}\right) = \]
\begin{equation} 
\left( \begin{array}{r r r r r r }
1 & 0.422703 & -0.006118 & 0.000023 & 0.000000 & 0.000000 \\
0 & 1 & -0.211351 & 0.070063 & -0.025553 & 0.009775 \\
0 & 0 & 1 & -0.845406 &  0.548271 & -0.317675 \\
0 & 0 & 0 & 1 & -1.479461 & 1.428504 \\
0 & 0 & 0 & 0 & 1 &  -2.113515 \\
0 & 0 & 0 & 0 & 0 & 1  \label{e4.8} \\  
\end{array} \right)
\end{equation} 

The transformation of the Taylor series in $K$ into Taylor series in $N$ is now achieved by applying eqn.s \eqref{e4.6}, \eqref{e4.7} and \eqref{e4.8} to the coefficient vectors in table 1.  

\subsection{Taylor coefficients }

Again we collect the Taylor coefficients, this time of $X^{-1}(N)$, $b(N)$, $B(N)$ and $a(N)$, in a table. For reasons that will soon become clear, we start the table with the coefficients of a function we have encountered in section 1.5: $c \equiv b^{-1/3}X^{-4/3}$. 

\begin{table}[H]
\begin{tabular}{ |l|r|r|r|r|r|}
\hline
$f$  & $c$ & $X^{-1}$ & $b$ & $B$ & $a$ \\
\hline
$\alpha$  & $-$2/3 & $-$2/3 & $-$2/3 & 1/3 & $-$2/3 \\
$\tilde f_0$ & 0.337821 & 0.337821 & 0.337821 & 1.013463 & 0.337821 \\
$\tilde f_1$ & $-$0.121969 & $-$0.234576 & $-$0.572397 & $-$0.429297 & 1.454528 \\
$\tilde f_2$ & $-$0.022859 & $-$0.019738 & 0.214837 & 0.092072 & $-$0.214459 \\ 
$\tilde f_3$ & $-$0.011826 & $-$0.011507 & 0.008230 & 0.002471 & 0.008229\\
$\tilde f_4$ & $-$0.007633 & $-$0.007524 & 0.003983 &  0.000907 & 0.001520\\ 
$\tilde f_5$ & $-$0.005504 & $-$0.005410 & 0.002114 & 0.000445 & 0.000249\\
\hline
\end{tabular}
\caption{Coefficients of the Taylor series $f=N^\alpha (\tilde f_0+\tilde f_1 N+\tilde f_2 N^2+\tilde f_3 N^3+ \ldots)$ for selected functions $f$.}
\end{table}

The three relations collected in section 1.5 can be restated as recursive relations between the Taylor-coefficients of $X^{-1}(N)$, $b(N)$, $B(N)$ and $a(N)$:
\begin{itemize}
\item Eqn \eqref{e1.2}\enskip :  $b=X^{-1}(1-N)$ \quad \qquad $\Rightarrow$  \quad $b_n = X^{(-1)}_n - X^{(-1)}_{n-1}$ \\
\item Eqn \eqref{e1.7}\enskip:  $B = \int_0 dN b$  \quad \qquad  \qquad  $\Rightarrow$ \quad $B_n=b_n/(n+1/3)$ \\
\item Eqn \eqref{e1.16}:  $a =7B/3+b(1-N)$ \enskip\;$\Rightarrow$ \quad $a_n = 7B_{n-1}/3+b_n-b_{n-1}$ \\
\end{itemize}

The reader may check that the coefficients in table 3 fulfill all these relations. In particular, the series for $X^{-1}$, $a$ and $b$ have the same first term $0.3378N^{-2/3}$.  

\subsection{Convergence }

As we did for the $f(K)$-series, we check the convergence of the $f(N)$-series for the neutral atom, $N=1$.  

\begin{table}[H]
\begin{tabular}{ |l|r|r|r|r|r|}
\hline
$f$ & $c$ & $X^{-1}$ & $b$ & $B$ & $a$ \\
\hline
$S_0$ & 0.337821 & 0.337821 & 0.337821 & 1.013462 & 0.337821  \\
$\ldots$ & \ldots & \ldots & \ldots & \ldots  & \ldots\\
$S_3$ & 0.181166 & 0.071998 & $-$0.011507 & 0.678709 & 1.586119 \\
$S_4$ & 0.173532 & 0.064474 & $-$0.007524 & 0.679617 & 1.587639 \\
$S_5$ & 0.168028 & 0.059063 & $-$0.005410 & 0.680063 & 1.587889 \\
\hline
$S_\infty$ & 0.0977 & 0 & 0 & 0.680601 & 1.588071 \\
\hline
\end{tabular}
\caption{Partials sums  $S_n=N^\alpha (\tilde f_0+ \tilde f_1 N+\tilde f_2 N^2+\ldots +\tilde f_n N^n )$ for $N=1$ (neutral atom) and selected functions $f$.}
\end{table}

The convergence is rather slow for $c(N)$ and $X^{-1}(N)$, fair for $b(N)$, and excellent for $B(N)$ und $a(N)$. Eliminating the parameter $K$ brought about an unexpected improvement in convergence! 

\section{Improved series for $X^{-1}(N)$, $b(N)$, $B(N)$ and $a(N)$}

When we want to describe a quantity $f(N)$ by a series, we actually have a lot of choices. We have encountered this situation in section 2, when we calculated Taylor series for $\eta$ and $\eta^{1/2}$, or for $\xi$ and $\xi^{-1}$. Both series contain the same information, and the question arises: which series converges faster? We leave this question unanswered for the $K$-series, and turn our attention to the (more important) $N$-series. In the last chapter, we had a loose hierarchy of the functions of interest:  $X^{-1}(N) \rightarrow b(N) \rightarrow B(N) \rightarrow a(N)$. The Taylor coefficients of each function can be calculated from the Taylor coefficients of the previous functions, with $X^{-1}(N)$ being the fundamental function. We now declare 
\begin{equation}
b^{-1/3}X^{-4/3} \equiv c(N)= \sum_{n=0}^\infty c_n N^{n-2/3} \label{e5.1}
\end{equation}
to be our fundamental function. Its Taylor coefficients $c_n$ are listed in table 3. 

\begin{figure}[!h]
\begin{center}
\includegraphics[trim = 0pt 0pt 0pt 0pt, scale=1]{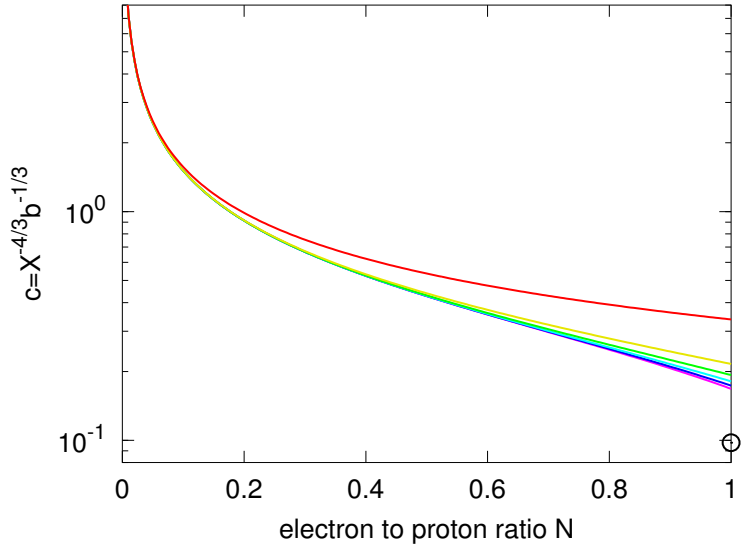}
\caption{Successive approximations of the fundamental function $c(N)$ with the Taylor series \eqref{e5.1}. The color of the orders is red (0), yellow (1), green (2), cyan (3), blue (4), magenta (5). For $N=1$, the series converges towards 0.0977 (o).}
\end{center}
\end{figure}

Our quantities of interest $X(N)$, $b(N)$, $B(N)$ and $a(N)$ are obtained as follows: 

The reciprocal ion radius is (see eqn.s \eqref{e1.2} and \eqref{e5.1})
\begin{equation}
X^{-1}(N)=(1-N)^{1/3}c(N)=\sum_{n=0}^\infty c_n N^{n-2/3}(1-N)^{1/3} \label{e5.2}
\end{equation}

The ionization potential  is (see eqn.s \eqref{e1.2} and \eqref{e5.1})
\begin{equation}
b(N)=(1-N)^{4/3}c(N)=\sum_{n=0}^\infty c_n N^{n-2/3}(1-N)^{4/3} \label{e5.3}
\end{equation}

The binding energy $B(N)$ is the primitive of $b(N)$ (see eqn.s \eqref{e1.7} and \eqref{e5.3}): 
\begin{equation}
B(N)=\int_0 dN (1-N)^{4/3}c(N)=\sum_{n=0}^\infty c_n\tilde B(N;n+1/3,7/3) \label{e5.4} 
\end{equation}
where $\tilde B$ is Euler's incomplete Betafunction.

The initial slope of the screening function is (see eqn.s \eqref{e1.2}, \eqref{e1.18} and \eqref{e5.1}):
\begin{align}
a(N)=\int (1-N)^{7/3}dc =& c_0\left[ N^{-2/3}(1-N)^{7/3} +\tfrac{7}{3}\tilde{B}(N;1/3,7/3) \right]+ \nonumber \\
&\sum_{n=0}^\infty c_n (n-2/3)\tilde B(N;n-2/3,10/3) \label{e5.5}
\end{align}
Because of its divergence, the first term of $a(N)$ needs a separate analysis. 

Why is $c(N)$ a smarter fundamental function than $X^{-1}(N)$? Well, the latter function tends towards $0$ for $N \rightarrow 1$. The Taylor series of $X^{-1}(N)$ cannot reproduce this behaviour well, in the sense that the relative error becomes infinite. On the other hand, the function $c(N)$ has a non-zero value for $N=1$: $c(N=1) = 0.0977$ (see Appendix B), and therefore can be reproduced with a Taylor series with finite relative error. As a consequence, the above formulas for $X(N)$, $b(N)$, $B(N)$ and $a(N)$ all have the correct behaviour for $N \rightarrow 1$  in-built! 

By way of example, let's check this argument for the reciprocal ion radius $X^{-1}$. In fig.s (6) and (7), we compare the convergence of the Taylor series (see table 3) and of the improved series (see eqn \eqref{e5.2} and table 3). For $N \rightarrow 1$, the convergence of the improved series is indeed much better.

\begin{figure}[ht]
\begin{minipage}[b]{0.47 \linewidth}
\centering
\includegraphics[trim = 0pt 0pt 18pt 0pt, scale=1 ]{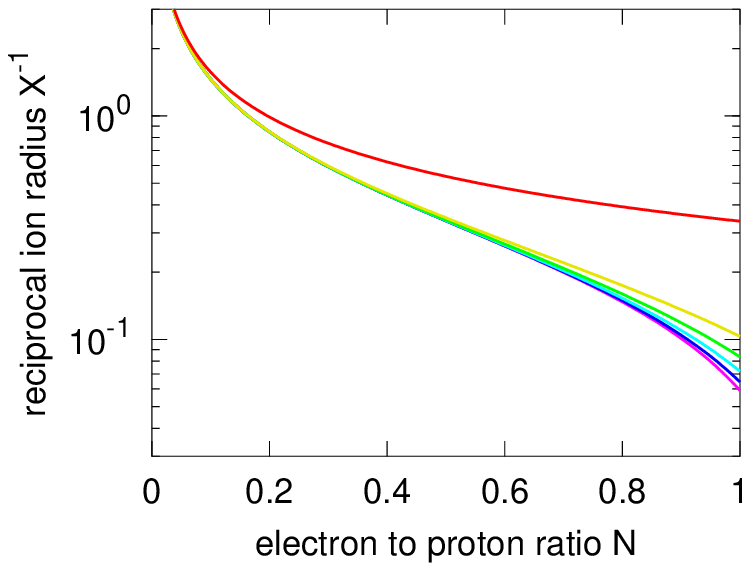}
\caption{Successive approximations of the reciprocal ion radius $X^{-1}(N)$ with the Taylor series given in table 3. }
\end{minipage}
\hspace{0.04\linewidth}
\begin{minipage}[b]{0.47\linewidth}
\centering
\includegraphics[trim = 0pt 0pt 18pt 0pt, scale=1]{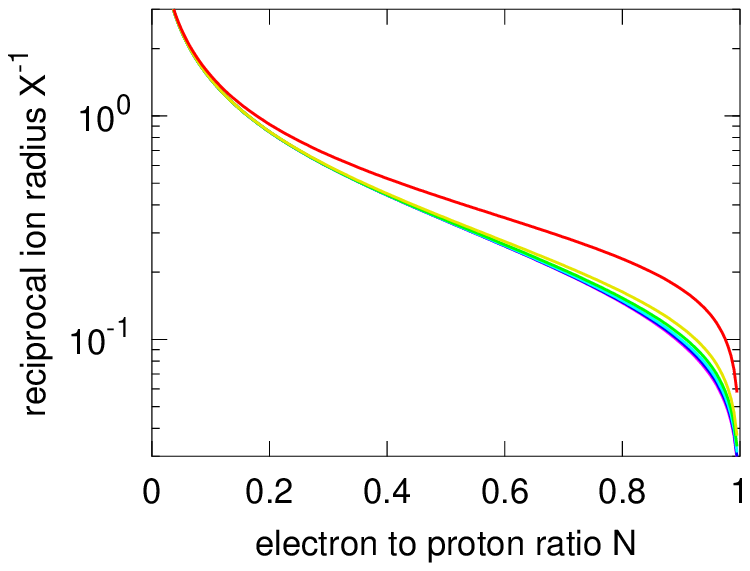}
\caption{Successive approximations of the reciprocal ion radius $X^{-1}(N)$ with the improved series \eqref{e5.2}.}
\end{minipage}
\end{figure}

\newpage
As a further test, let us compare the convergence of the Taylor series
(see tables 3 and 4)
\begin{equation*}
%a(N=1)=\tfrac{7}{3}\sum_{n=0}^\infty B_n \qquad (i) \qquad \qquad 
a(N=1)=\sum_{n=0}^\infty a_n \qquad (i) 
\end{equation*}
and of the improved series (see eqn\eqref{e5.5} and table 3)
\begin{equation*}
a(N=1)=\tfrac{7}{3}\sum_0^\infty c_n\tilde B(n+1/3,7/3)  \qquad (ii) 
\end{equation*}
where $\tilde B$ is now Euler's ''complete'' Betafunction, and where we have used the identity $p\tilde B(p,q+1)=q\tilde B(p+1,q)$.

\begin{tabular}{ |l|rrrrrr|r|}
\hline
$a$  & $a_0$ & $a_1$ & $a_2$ & $a_3$   & $a_4$ & $a_5$ &  \\
\hline
(i) & 0.337821 & 1.454528 & -0.214459 & 0.008229 & 0.001520 & 0.000249 & \\
(ii) & 1.671061 & -0.075416 & -0.005140 & -0.001330 & -0.000505 & -0.000237 & \\
\hline
$a$  & $S_0$ & $S_1$ & $S_2$ & $S_3$   & $S_4$ & $S_5$ & $S_\infty$\\
\hline
%(i) & 2.364747 & 1.363052 & 1.577888 & 1.583656 & 1.585774 & 1.586814 & 1.588071 \\
(i) & 0.337821 & 1.792349 & 1.577890 & 1.586119 & 1.587639 & 1.587889 & 1.588071 \\
(ii) & 1.671061 & 1.595645 & 1.590505 & 1.589176 & 1.588671 & 1.588434 &1.588071 \\
\hline
\end{tabular}

The outcome of the comparison is less clearcut here. Initially, the improved series $(ii)$ approaches the final value much faster than the Taylor series. However, as more terms are added, the convergence of the two series becomes similar, and series $(i)$ even seems to have the edge.

\section{Summing up and plots}

In this section we collect the formulas for the ion radius $X(N)$, the ionization potential $b(N)$, the electronic binding energy $B(N)$ and the initial slope of the screening function $a(N)$, where the independent variable $N$ is the electron to proton ratio of the ion. The fundamental quantity is $c \equiv b^{-1/3}X^{-4/3}$ with the Taylor  series 
\begin{align*} 
c(N)  =\sum_{n=0}^\infty c_n N^{n-2/3} &= 0.337821N^{-2/3}-0.121969N^{1/3}-0.022859N^{4/3} \\
& -0.011826N^{7/3}-0.007633N^{10/3} -0.005504N^{13/3}
\end{align*}  
The ``improved series'' for $X^{-1}(N)$, $b(N)$, $B(N)$ and $a(N)$ are all expressed in terms of the Taylor series of $c(N)$.
We plot the partial sums of the ``improved series'' of the four quantities. The color of the orders is red (0), yellow (1), green (2), cyan (3), blue (4), magenta (5). The convergence is excellent in all cases. 

\newpage
\subsection{The radius of the TF ion}

The ion radius $X$ in units of $a_B/1.1295 Z^{1/3}$ is given by  $X^{-1}(N)=(1-N)^{1/3}c(N)$.

\begin{figure}[h!]
\begin{center}
\includegraphics[trim = 0pt 0pt 0pt 0pt, scale=1]{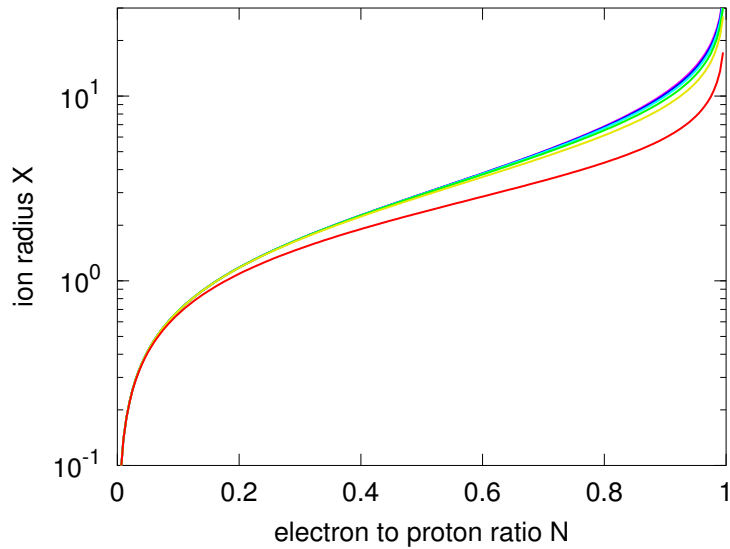}
\caption{Successive approximations of the ion radius $X$.}
\end{center}
\end{figure}

\subsection{The ionization potential of the TF ion}

The ionization potential in units of $2.2590 Z^{4/3} Ry$  is $b(N)=(1-N)^{4/3}c(N)$.

\begin{figure}[!h]
\begin{center}
\includegraphics[trim = 0pt 0pt 0pt 0pt, scale=1]{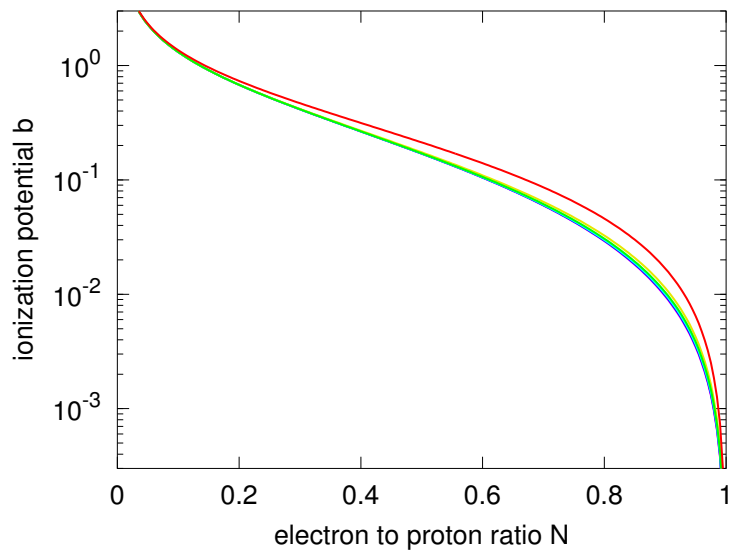}
\caption{Successive approximations of the ionization potental $b$.}
\end{center}
\end{figure}

\newpage
\subsection{The electronic binding energy of the TF ion}

The electronic binding energy in units of $2.2590 Z^{7/3} Ry$ is $B(N)=\sum_0^\infty c_n\tilde B(N;n+1/3,7/3)$, where $\tilde B$ is Euler's incomplete Betafunction. 

\begin{figure}[!h]
\begin{center}
\includegraphics[trim = 0pt 0pt 0pt 0pt, scale=1]{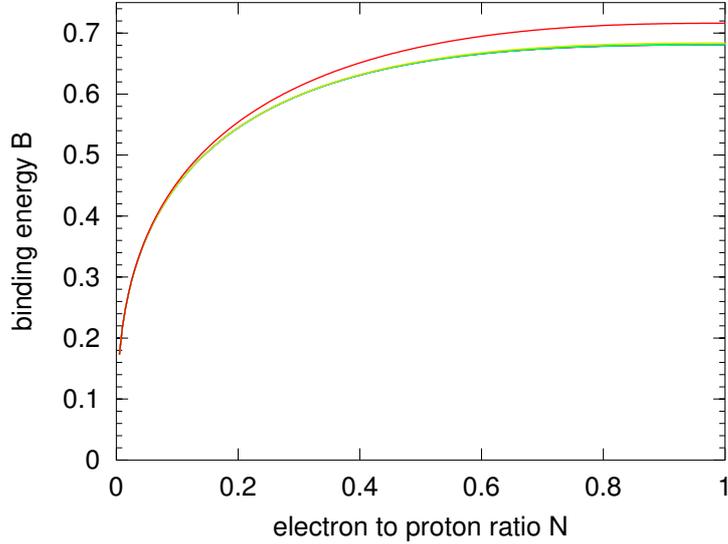}
\caption{Successive approximations of the electronic binding energy $B$.}
\end{center}
\end{figure}

\subsection{The initial slope of the screening function}

The initial slope of the screening function is $a(N)=c_0\left[ N^{-2/3}(1-N)^{7/3} +\tfrac{7}{3}\tilde{B}(N;1/3,7/3) \right] +\sum_{n=1}^\infty c_n (n-2/3)\tilde B(N;n-2/3,10/3)$. In the following figure, the related quantity $K=2/a^{3/2}$ is plotted. 

\begin{figure}[!h]
\begin{center}
\includegraphics[trim = 0pt 0pt 0pt 0pt, scale=1]{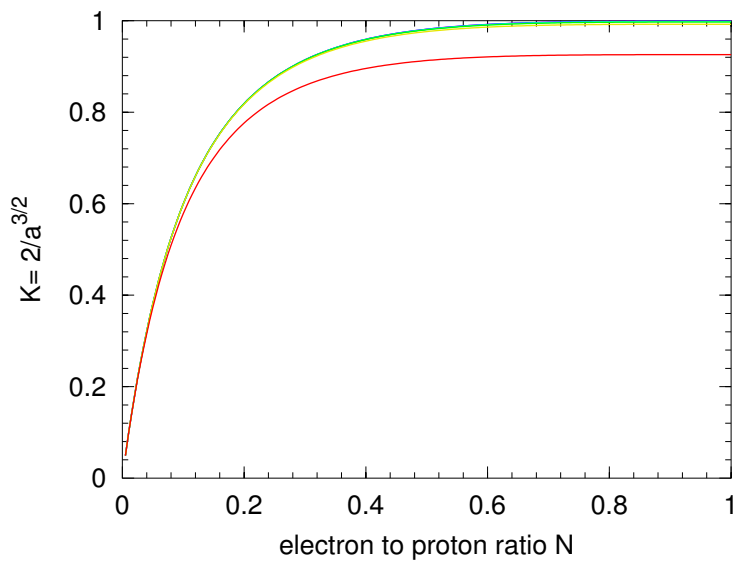}
\caption{Successive approximations of the parameter $K=2/a^{3/2}$.}
\end{center}
\end{figure}

\newpage
\section{A note about the screening function}

By means of the integration scheme described in section 2, we obtained a Taylor series for the ion radius $X(K)$. By applying the linear transformation described in section 4, we obtained the corresponding series for $X(N)$. Actually, in section 2 we obtained a series in $K$ for the whole inverse screening function $x(\chi;K)$; the ion radius $X$ is simply the value of the inverse screening function for $\chi=0$. So, why not apply the transformation from $K$- to $N$-series to the whole inverse screening function? When the author did this, he found to his surprise (and dismay) that the series  
\begin{equation}
x(\chi;N)=N^{2/3}\left(x_0(\chi)+ x_1(\chi)N+ x_2(\chi)N^2+ x_3(\chi)N^3+ \ldots\right)
\end{equation}
diverges for all values of $\chi$ except for the border values $\chi=1$ (ion center) and $\chi=0$ (ion edge). When it comes to calculating the screening function of the TF ion, this article has therefore not much to contribute.

\newpage

\appendix

\section{Appendix: Integral formulas}

Eqn \eqref{e1.5} for the binding energy $B$ 
contains integrals of the type 
\begin{equation}
I_0(\beta)=\int_0^X dx x^{\beta-1/2} \chi^{3/2} =\int_a^b d\psi x^\beta 
\end{equation}
and
\begin{equation}
I_1(\beta)=\int_0^X dx x^{\beta-1/2}\chi^{5/2}  =\int_a^b d\psi x^\beta \chi
\end{equation}
where
\begin{equation*} 
 \psi \equiv -\frac{d\chi}{dx} 
\end{equation*}
and the TF differential eq. 
\begin{equation*} 
-\frac{d\psi}{dx}= \frac{\chi^{3/2}}{ x^{1/2}} 
\end{equation*}
is assumed to hold.

\subsection*{Evaluation of the integrals}

The first integral is 
\begin{equation*}
I_0(\beta)=\int_a^b d\psi x^\beta 
\end{equation*}

For $\beta=0$, integration by parts gives
\begin{equation}
I_0(0)=a-b  \label{eA.1}
\end{equation}

For $\beta>0$, integration by parts gives
\begin{equation*}
I_0(\beta)=\beta \int_0^1 d\chi x^{\beta-1}-bX^\beta  
\end{equation*}
The case $\beta=1$ is readily evaluated: 
\begin{equation}
I_0(1)=1-bX \label{eA.2}
\end{equation}

The second integral is
\begin{equation*}
I_1(\beta)=\int_b^a d\psi x^\beta \chi
\end{equation*}

For $\beta=0$, integration by parts gives
\begin{equation*}
I_1(0)=a-\int_0^1 d\chi \psi 
\end{equation*}
Setting $d\chi=-\psi dx$ and integrating by parts gives
\begin{equation*}
I_1(0)=a-b^2X -2 \int_b^a d\psi \psi x
\end{equation*}
From \eqref{e2.10}, $\psi d \psi = x^{-1/2} \chi^{3/2} d\chi$, and therefore
\begin{equation*}
I_1(0)=a-b^2X -2 \int_0^1 d\chi x^{1/2} \chi^{3/2}
\end{equation*}
Integrating the last integral one more time by parts gives
\begin{equation*}
I_1(0)=a-b^2X -\frac{2}{5} \int_0^X dx x^{-1/2} \chi^{5/2}
\end{equation*}
But the last integral is again $I_1(0)$ ! Bringing this term to the LHS, we finally obtain
\begin{equation}
I_1(0)=\frac{5}{7}(a-b^2X) \label{eA.3}
\end{equation}

For $\beta>0$, integration by parts gives
\begin{equation*}
I_1(\beta)=\beta \int_0^X dx \psi \chi x^{\beta-1} - \int_0^1 d\chi x^\beta \psi
\end{equation*}
The rest of the calculation is analogous to the case $\beta=0$. The final result is 
\begin{equation*}
I_1(\beta)=\frac{5}{9\beta+7} \left( (\beta+1)\beta \int_0^1 d\chi \chi x^{\beta-1} - b^2 X^{\beta+1} \right)
\end{equation*}
 
The case $\beta=1$ is readily evaluated: 
\begin{equation}
I_1(1) =\frac{5}{16} \left( 1 - b^2 X^2 \right) \label{eA.4}
\end{equation}

\subsection*{Summary and physical meaning}

We have evaluated four integrals. 

The first integral \eqref{eA.1} is the electrostatic binding energy $B_{pe}$ between nucleus and electrons:    
\begin{equation*}
B_{pe}=I_0(0)=\int_0^X dx x^2 \frac{\rho}{x} =a-b  
\end{equation*}

The second integral \eqref{eA.2} is the $e$:$p$-ratio $N$:
\begin{equation*}
N=I_0(1)=\int_0^X dx x^2 \rho = 1-bX  
\end{equation*}

The third integral \eqref{eA.3} is the kinetic energy of the electrons $T$ times a factor:
\begin{equation*}
\frac{5}{3}T=I_1(0)=\int_0^X dx x^2 \rho^{5/3}=\frac{5}{7}(a-b^2X)
\end{equation*}

The binding energy $B$, eqn \eqref{e1.5}, contains contributions from these three integrals: 
\begin{equation*}
 B=\frac{1}{2}I_0(0)+\frac{b}{2}I_0(1)-\frac{1}{10}I_1(0)=\frac{3}{7}(a-b^2 X)
\end{equation*}

The last two equations are in agreement with the virial theorem: the kinetic energy $T$ of a system of electrons trapped in the Coulomb potential of a nucleus equals the electronic binding energy $B$.  

The author can't recognise a physical meaning in the fourth integral \eqref{eA.4}. By combining  it with eqn \eqref{e1.2}: $N=1-bX$, we obtain the curious relation
\begin{equation*}
I_1(1)=\int_0^X dx x^2 \rho \chi  =\frac{5}{16} N(2-N)
\end{equation*}

\section{Appendix: The limiting value of c}

The two quantities

\begin{minipage}[b]{0.45\linewidth} 
\begin{equation} 
C \equiv X^{4/5}b^{1/5}
\end{equation} 
\end{minipage}
\begin{minipage}[b]{0.45\linewidth} 
\begin{equation} 
c \equiv X^{-4/3}b^{-1/3} = C^{-5/3} 
\end{equation} 
\end{minipage} 

contain the factors $b$ and $X$ raised to a power. $X$ tends to $\infty$ for $N\rightarrow 1$, and $b$ tends to $0$ for $N\rightarrow 1$. The following calculation shows that $C$ and $c$ assume finite and non-zero values for $N\rightarrow 1$.    

Let us formally integrate eqn.s \eqref{e2.9} and \eqref{e2.10}, this time starting at $\chi = 0$: 

\begin{minipage}[b]{0.45\linewidth} 
\begin{equation*} 
x=  X-\int_0 \frac{d\chi}{\psi} 
\end{equation*} 
\end{minipage} 
\begin{minipage}[b]{0.45\linewidth} 
\begin{equation*} 
\psi^2 =  b^2+2\int_0 d\chi \frac{\chi^{3/2}}{x^{1/2}} 
\end{equation*} 
\end{minipage}

The two integral equations can be normalized by substituting

\begin{minipage}[b]{0.45\linewidth} 
\begin{equation*} 
\xi \equiv \frac{x}{X}
\end{equation*} 
\end{minipage} 
\begin{minipage}[b]{0.45\linewidth} 
\begin{equation*} 
\eta \equiv \left(\frac{\psi}{b}\right)^2 
\end{equation*} 
\end{minipage}
\begin{minipage}[b]{0.45\linewidth} 
\begin{equation*} 
t  \equiv \frac{\chi}{X^{1/5}b^{4/5}} 
\end{equation*} 
\end{minipage} 
\begin{minipage}[b]{0.45\linewidth} 
\begin{equation*} 
T  \equiv \frac{1}{X^{1/5}b^{4/5}}
\end{equation*} 
\end{minipage} 

The result is 

\begin{minipage}[b]{0.45\linewidth} 
\begin{equation} 
\xi  =1-\frac{1}{C} \int_0 \frac{dt}{\eta^{1/2}}
\end{equation} 
\end{minipage} 
\begin{minipage}[b]{0.45\linewidth} 
\begin{equation} 
\eta  = 1+2 \int_0 dt \frac{t^{3/2}}{\xi^{1/2}} 
\end{equation} 
\end{minipage} 

$\quad$ \\
The quantity $C$ is a function of the $e$:$p$-ratio $N$, and thus parametrizes the solutions $\xi(t)$ and $\eta(t)$. The equations $x(\chi=1)=0$ and $\psi^2(\chi=1)=a^2$ translate to $\xi(t=T)=0$ and $\eta(t=T)=(a/b)^2$. It follows that

\begin{minipage}[b]{0.45\linewidth} 
\begin{equation*} 
C = \int_0^T \frac{dt}{\eta^{1/2}}
\end{equation*} 
\end{minipage} 
\begin{minipage}[b]{0.45\linewidth} 
\begin{equation*} 
\left( \frac{a}{b}  \right) ^2 = 1+2 \int_0^T dt \frac{t^{3/2}}{\xi^{1/2}} 
\end{equation*} 
\end{minipage} 

We are only interested in the limes $N \rightarrow 1$, where $T \rightarrow \infty$. The left eqn. then becomes. 
\begin{equation} 
C = \int_0^\infty \frac{dt}{\eta^{1/2}} 
\end{equation} 

Eqn.s (B.3), (B.4), (B.5) can be solved iteratively on the computer, beginning with $C=\infty \rightarrow \xi=1$. The result is 

\begin{minipage}[b]{0.45\linewidth} 
\begin{equation} 
C=4.03623
\end{equation} 
\end{minipage} 
\begin{minipage}[b]{0.45\linewidth}
\begin{equation} 
c=0.097733
\end{equation} 
\end{minipage} 

\end{flushleft} 
\end{document}